\documentclass[11pt]{article}
\usepackage{epsfig,a4}
\def\be{\begin{equation}}
\def\ee{\end{equation}}
\def\bea{\begin{eqnarray}}
\def\eea{\end{eqnarray}}

\def\C{{\rm\kern.24em
    \vrule width.02em height1.4ex depth-.05ex
    \kern-.26em C}}
\def\N{{\rm I\kern-.18em N}}
\def\R{{\rm I\kern-.21em R}}
\def\Z{{\rm\kern.26em
    \vrule width.02em height0.5ex depth 0ex
    \kern.04em
    \vrule width.02em height1.47ex depth-1ex
    \kern-.34em Z}}
\def\d{{\rm\kern.22em
    \vrule width.02em height1.0ex depth0ex
    \kern-.24em d}}
\def\nn{\nonumber}

\def\rh{\rho}
\def\ta{\tau}
\def\ps{\psi}
\def\Ga{\Gamma}
\def\ga{\gamma}
\def\rh{\rho}
\def\ep{\epsilon}
\def\de{\delta}
\def\al{\alpha}

\def\cD{{\cal D}}
\def\cS{{\cal S}}
\def\bB{\mbox{\bf B}}
\def\bV{\mbox{\bf V}}
\newcommand{\vDelta}{\vec\Delta}
\newcommand{\vdelta}{\vec\delta}
\newcommand{\lbq}[1]{\label{#1} \enq}

\newcommand{\beq}{\begin{equation}}
\newcommand{\enq}{\end{equation}}
\newcommand{\beqa}{\begin{eqnarray}}
\newcommand{\beqast}{\begin{eqnarray*}}
\newcommand{\enqa}{\end{eqnarray}}
\newcommand{\enqast}{\end{eqnarray*}}
\newcommand{\lb}{\label}
\newcommand{\rf}{\ref}

\newcommand{\pa}{\partial}

\def\O{{\rm\kern.24em
    \vrule width.02em height1.4ex depth-.05ex
    \kern-.26em O}}

\newdimen\picraise
\newcommand\picbox[1]
{
  \setbox0=\hbox{\input{#1}}
  \picraise=-0.5\ht0 
  \advance\picraise by 0.5\dp0
  \advance\picraise by 3pt     
  \hbox{\raise\picraise \box0}
}
\newdimen\picraiset
\newcommand\picding[1]
{
  \setbox0=\hbox{\input{#1}}
  \picraiset=-0.5\ht0 
  \advance\picraiset by 0.5\dp0
  \advance\picraiset by -3pt  
  \hbox{\raise\picraiset \box0}
}
\newdimen\picraisehallo
\newcommand\pichallo[2]
{
  \setbox0=\hbox{\input{#1}}
  \picraisehallo=-0.5\ht0 
  \advance\picraisehallo by 0.5\dp0
  \advance\picraisehallo by -#2pt  
  \hbox{\raise\picraisehallo \box0}
}
\begin{document}
\begin{titlepage}
\begin{flushright}
HD-THEP-01-45\\
hep-ph/0201294
\\
\end{flushright}
\vfill
\begin{center}
\boldmath
{\Large{\bf The Odderon in High Energy Elastic $pp$ Scattering $^*$}}
\unboldmath
\end{center}
\vspace{1.2cm}
\begin{center}
{\bf \large
Hans G\"unter Dosch $^a$, Carlo Ewerz $^b$, Volker Schatz $^c$
}
\end{center}
\vspace{.2cm}
\begin{center}
{\sl
Institut f\"ur Theoretische Physik, Universit\"at Heidelberg\\
Philosophenweg 16, D-69120 Heidelberg, Germany}
\end{center}                                                                 
\vfill
\begin{abstract}
We study the Odderon contribution to 
elastic $pp$ and $p\bar{p}$ scattering at high energies. 
Different models for the Odderon--proton coupling 
are considered and their effects on the differential 
cross section in the dip region are investigated. 
We use a Regge fit by Donnachie and Landshoff as a framework 
and replace its Odderon contribution by the different 
models. 
We consider two models for the Odderon--proton impact 
factor proposed by Fukugita and Kwieci\'nski and 
by Levin and Ryskin. In addition we construct a 
geometric model of the proton which allows us to 
put limits on the size of a possible diquark 
cluster in the proton. 
All models are able to describe the data well. 
The two models for the impact factor require 
the strong coupling constant to be fixed rather precisely. 
In the geometric model a relatively small diquark 
size is required to describe the data. 

%
%
%
\vfill
\end{abstract}
\vspace{5em}
\hrule width 5.cm
\vspace*{.5em}
{\small \noindent 
$^*$Work supported in part by the EU Fourth Framework Programme
`Training and Mobility of Researchers', Network `Quantum Chromodynamics
and the Deep Structure of Elementary Particles',
contract FMRX-CT98-0194 (DG 12 - MIHT).\\[.2cm]
$^a$ email: H.G.Dosch@thphys.uni-heidelberg.de \\
$^b$ email: C.Ewerz@thphys.uni-heidelberg.de \\
$^c$ email: V.Schatz@thphys.uni-heidelberg.de 
}
\end{titlepage}

\section{Introduction}
\label{intro}
The Odderon is an interesting but elusive object. 
Its history goes back to 1973 when the possible contribution 
of an exchange carrying negative $C$ parity 
to very high energy collisions was first discussed \cite{Lukaszuk:1973nt}. 
The leading contribution to hadronic scattering processes 
is in general well described by the exchange of a Pomeron 
with intercept $\alpha \simeq 1.09$, resulting in 
slowly rising cross sections, $\sigma \sim s^{\alpha-1}$ 
\cite{Groom:in}. 
The Pomeron carries vacuum quantum numbers and therefore 
leads to a high energy behaviour of hadronic 
cross sections that is equal for $pp$ and $p\bar{p}$ 
scattering. In lowest 
order in QCD the Pomeron can be identified with the exchange of 
two gluons in a colour singlet state. The Odderon is the 
$C=-1$ partner of the Pomeron. In lowest order it can be understood 
as the exchange of three gluons in a symmetric colour 
singlet state. As in the case of the Pomeron, the Odderon exchange 
gives a contribution to the cross section that behaves like a power 
of the energy, $s^{\alpha_{\cal O}-1}$. 
The Odderon intercept $\alpha_{\cal O}$ 
is expected to be close to one --- in contrast to the intercept of 
$C=-1$ reggeon exchanges which is around $0.5$. 
For a review of the historic roots of the Odderon and some relevant 
references we refer the reader to \cite{Braun:1998fs}. 

\begin{figure}[htbp]
\begin{center}
\input{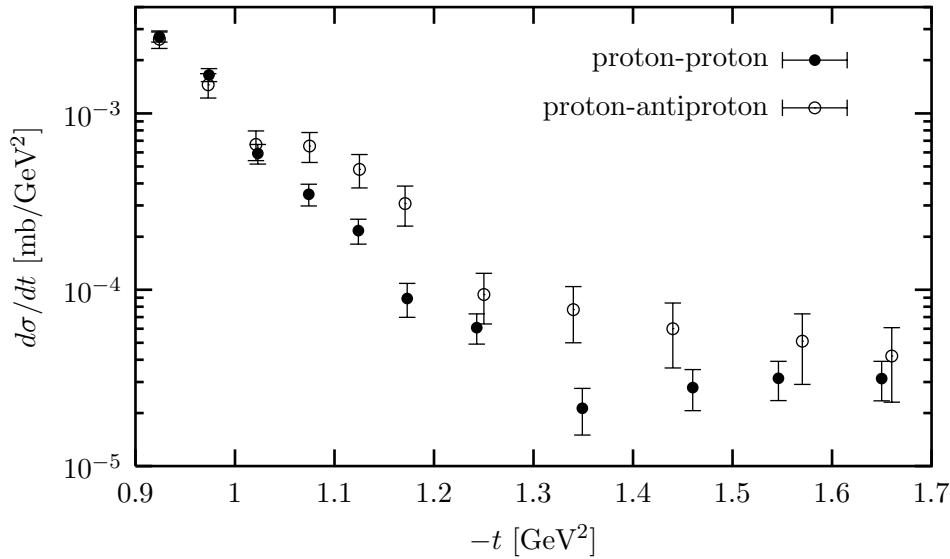}
\end{center}
\caption{Differential cross section for elastic $pp$ and $p\bar{p}$ 
scattering in the dip region for $\sqrt{s}=53\,\mbox{GeV}$; 
data taken from 
\protect\cite{Breakstone:1985pe}}
\label{fig:dipdiff}
\end{figure}

The experimental evidence for existence of an Odderon, however, 
remains rather scarce. For a long time the Odderon search concentrated 
on observing a difference between 
the cross sections for $pp$ and $p\bar{p}$ scattering 
at high energies. The Odderon causes such a difference because it 
carries negative $C$ parity and thus gives opposite contributions 
to these cross sections. For an intercept larger than one this 
difference should in fact increase with energy and give a visible 
effect. But recent perturbative results indicate that the intercept 
is smaller than or equal to one such that the difference does not 
increase. The experimental data actually disfavour a sizable effect 
of the Odderon in this difference. 
More sensitive to the Odderon exchange than the cross section 
is the ratio of the real to the imaginary part of the scattering 
amplitude in the forward direction. But also there no indication 
of an Odderon exchange has been found \cite{Augier:1993sz}. 
To date the only evidence for the existence of the Odderon 
is found in the $t$-dependence of $pp$ and $p\bar{p}$ 
elastic cross sections at high energy. The $pp$ data show 
a characteristic dip at $|t|\simeq 1.3\, \mbox{GeV}^2$, 
whereas the $p\bar{p}$ data only flatten off at that $t$. Unfortunately, 
there are only few $p\bar{p}$ data available 
\cite{Breakstone:1985pe,Erhan:1985mv}. 
Figure \ref{fig:dipdiff} shows the data 
\cite{Breakstone:1985pe} in the relevant region of the dip. 
A good description of all available data 
\cite{Breakstone:1985pe}--\cite{Amaldi:1980kd} for elastic $pp$ 
scattering was given by Donnachie and Landshoff \cite{Donnachie:1984hf}. 
In this description (see also section \ref{dlsection} below) the 
presence or absence of the dip originates from the exchange 
of the Odderon. Also the behaviour of the elastic cross sections 
at large $t$ is well described by Odderon exchange. 
A number of aspects of elastic $pp$ and $p\bar{p}$ scattering 
has been discussed in the literature particularly in the light 
of Odderon physics, see for example 
\cite{Gauron:1985gj}--\cite{Desgrolard:2000dn}. 
No other successful description of the data without an Odderon has 
been found so far. 

Recently there has been renewed interest in the Odderon 
which has especially concentrated on two areas. 
One of these areas is the perturbative treatment of the Odderon 
and in particular the determination of the Odderon 
intercept \cite{Gauron:1993dc}--\cite{Bartels:2000yt}. 
Perturbation theory can be applied to this problem 
if a scattering process involves a large momentum scale. 
The perturbative Odderon is described by the 
Bartels--Kwieci\'nski--Prasza{\l}owicz equation 
\cite{Bartels:1980pe,Kwiecinski:1980wb} 
which resums large logarithms of the center--of--mass 
energy. It  has the form of a Bethe--Salpeter type 
equation for the interaction of a system of three reggeized 
gluons in the $t$-channel. The interaction of the gluons 
induces a non--trivial energy dependence of the cross 
section, i.\,e.\ it leads to an intercept different from one. 
A number of interesting aspects of the perturbative Odderon 
has been studied in \cite{Lipatov:1993qn}--\cite{Derkachov:2001yn}. 
In \cite{Lipatov:1993yb,Faddeev:1994zg} it was shown that this system of 
three reggeized gluons is equivalent to an integrable model. 
Eventually the study of this system lead to the determination 
of its ground state energy \cite{Wosiek:1996bf,Janik:1998xj}. 
It was found to correspond to an intercept slightly below one. 
There exists, however, a special solution with intercept 
exactly equal to one \cite{Bartels:2000yt}. 
In \cite{Korchemsky:2001nx} even the whole energy spectrum 
of the perturbative Odderon was found. Another aspect of the 
perturbative Odderon is its r\^ole played in the theory of unitarity 
corrections to the perturbative (BFKL) Pomeron 
\cite{Kuraev:fs,Balitsky:ic}. 
This problem was addressed in \cite{Bartels:1999aw} where the 
perturbative Pomeron--Odderon--Odderon vertex was calculated. 

The other area on which interest has concentrated is to find 
more exclusive processes in which the Odderon contribution 
should be dominant. Some processes have been 
considered which can be calculated perturbatively. 
The most interesting among them is the 
diffractive photo-- or electroproduction 
of $\eta_c$ or other heavy pseudoscalar mesons at HERA 
\cite{Czyzewski:1997bv,Engel:1997cg,Bartels:2001hw}.
If these mesons have charge parity $+1$ Pomeron 
exchange cannot contribute to their photo-- or electroproduction 
\cite{Schafer:1992pq}. 
However, the corresponding cross sections are estimated to be rather 
small, in the range of several tens of pb or even lower. 
Much larger cross sections are expected for the diffractive 
production of light pseudoscalar and tensor mesons 
\cite{Kilian:1998ew,Ryskin:1998kt}. 
In this case the theoretical predictions require the use of 
nonperturbative methods and models 
\cite{Rueter:1998gj,Berger:1999ca,Berger:2000wt}. 
But recent measurements of pion photoproduction by the H1 collaboration 
did not show any signs of the Odderon \cite{Olsson:2001nm,Golling:2001ju}. 
The reasons for this failure of the prediction are not clear at the 
moment, and the presence of this process should in fact not be specific to 
the model assumptions. 

Recently it was proposed to investigate certain charge asymmetries 
in diffractive processes \cite{Brodsky:1999mz,Ivanov:2001zc}. 
These asymmetries arise from Pomeron--Odderon interference 
and are expected to provide a good chance of finding the Odderon at 
HERA. Another process of interest will be the quasidiffractive 
production of $\eta_c$ (or other pseudoscalar or 
tensor) mesons in collisions of real or 
virtual photons at a future linear collider like TESLA 
\cite{Motyka:1998kb}. An interesting process for the Odderon 
search is also double--diffractive production of vector 
mesons at Tevatron or at the LHC \cite{Schafer:1991na}. 

The apparent absence of the Odderon in the photoprodution of pions  
mentioned above is rather surprising. 
Its cause is an open question which clearly needs 
to be clarified. One obvious possibility is that for some reason the 
nonperturbative model used here does not work properly in this particular 
situation and that the cross section has thus been overestimated. 
This uncertainty can be excluded in perturbative situations. Therefore 
the investigation of perturbative processes involving the Odderon becomes 
even more important.  

A theoretical uncertainty that is inherent even in the perturbatively 
calculable processes like diffractive $\eta_c$ production 
is the exact form of the coupling of the Odderon to the proton, i.\,e.\ 
the Odderon--proton impact factor. 
A few very general facts about its structure are known, 
but some model assumptions always need to be made. 
But it is well known that the proton structure can in fact 
have very dramatic effects on this impact factor. In the 
extreme case that the proton would exhibit a quark--diquark 
structure with a pointlike diquark, for example, the impact 
factor even vanishes \cite{Zakharov:1989bh,Rueter:1996yb}. 
It was pointed out \cite{Rueter:1996yb} that even a diquark 
cluster of a size as large as $0.3\,\mbox{fm}$ could explain 
the experimental limit for the difference in the ratios of 
the real and imaginary part for $pp$ and $p\bar{p}$ forward 
scattering. 
It is the aim of the present paper to study the 
influence of the proton structure on the Odderon coupling 
and compare it with the available data for elastic $pp$ and 
$p\bar{p}$ scattering in the dip region. 
As a framework we use the fit by Donnachie and Landshoff 
\cite{Donnachie:1984hf}. We replace the Odderon--proton coupling 
used in that fit by a model for the proton structure which allows 
us to study the influence of a possible quark--diquark 
structure of the proton. The squared momentum transfer $t$ 
in the dip region appears to be large enough to make the use of 
the simple picture of perturbative three gluon exchange possible 
in which the Odderon has intercept one. A slow energy 
dependence of the Odderon due to logarithmic 
enhancements should not have a sizable effect in the 
restricted range of energies for which data are available. 
In a similar way we also test 
other Odderon--proton impact factors that have 
been used recently in diffractive $\eta_c$ production. 
The crucial point is that we are now able to see whether 
they are compatible with the only existing data
which clearly involve an Odderon exchange.  

The paper is organized as follows. In section 
\ref{dlsection} we briefly sketch the original 
Donnachie--Landshoff fit and describe in section \ref{impact} 
two different models for the Odderon--proton impact factor 
proposed in the literature. In subsection \ref{highenergyinpos} 
we present a geometrical model for the proton structure in 
position space. We show how this model can be implemented 
in a more general framework of high energy scattering 
in position space when applied to Odderon exchange. 
In section \ref{resultssec} we present the results for the 
differential cross section using these different models 
and confront them with data. 
Finally, we give a brief summary and conclusions in 
section \ref{conclusions}. 

\section{Odderon-proton coupling and proton structure}
\label{couplings}

In this section we discuss different ways in which the 
Odderon--proton coupling can be described. We start 
by giving a short description of the Donnachie--Landshoff (DL) 
fit and its making use of the Odderon. We then consider 
the perturbatively motivated description of the 
Odderon--proton coupling via impact factors. 
These impact factors are usually used in perturbative 
calculations and most naturally written in momentum 
space. We discuss two different models of the impact factors 
that have been proposed in the literature. 
Finally we turn to a geometrical picture of the proton 
as a three--quark system in which we can easily study 
the effects of a possible quark--diquark structure in the proton. 
Obviously this geometrical model of the proton is most 
conveniently formulated in position space. 
We therefore find it useful to give a description of 
Odderon exchange that works entirely in 
position space. We start from a general framework for 
high energy scattering and then derive a description of 
perturbative three--gluon exchange in position space. 
The use of a simple picture of the 
Odderon as an exchange of three 
perturbative gluons in the present paper is motivated 
by the fact that we are only considering 
the dip region of $pp$ scattering at around 
$|t| \simeq 1.3\, \mbox{GeV}^2$. This momentum 
transfer, however, is at the lower edge of the applicability 
of perturbation theory. A study of the dip region in a 
nonperturbative framework would therefore be 
very desirable. Although such a study is beyond 
the scope of the present paper we hope to pave 
the way for it by deriving the perturbative 
description of Odderon exchange in position space 
in a more general framework which can also be used 
to implement nonperturbative models. 

\subsection{The Donnachie-Landshoff fit}
\label{dlsection}

\begin{figure}[htb]
\begin{center}                                                               
\input{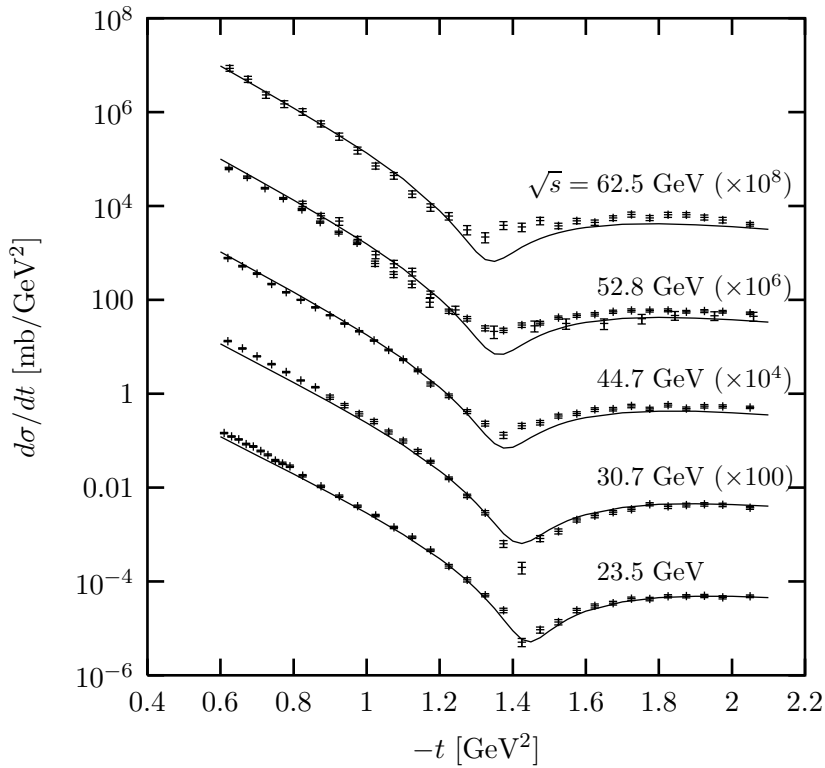}
\end{center}
\caption{The Donnachie-Landshoff fit for the differential elastic 
$pp$ cross section}
\label{fig:allsdl}
\end{figure}

A successful phenomenological description of 
all available $pp$ and $p\bar{p}$ elastic scattering 
data in the framework of Regge theory 
was given by Donnachie and Landshoff 
\cite{Donnachie:1984hf}. 
This description is based on a number of exchanges 
in the $t$-channel: Pomeron, reggeon, Odderon, 
double Pomeron, triple Pomeron, Pomeron plus two gluons, 
and reggeon plus Pomeron. 
These exchanges are explicitly given as contributions 
to the scattering amplitude $T(s,t)$. 
For later use we would like to single out the Odderon 
contribution $T^{\cal O}$ to that sum, 
\be
 T(s,t) = T^{\cal O}(s,t) + T^{\rm DL}(s,t) \,,
\ee
where $T^{DL}$ denotes all other contributions 
to the scattering amplitude, including the $C$-odd 
reggeon contribution. 
The differential cross section is then obtained from the 
scattering amplitude $T$ via 
\begin{equation}
\frac{d\sigma}{dt} = \frac{1}{16\,\pi\, s^2} \left|T(s,t)\right|^2
\,.
\label{dsigdtfromT}
\end{equation}
The different contributions to the scattering amplitude 
come with a number of parameters which have been fitted 
to all available data for elastic $pp$-scattering 
in \cite{Donnachie:1984hf}. The details and all parameters 
can be found in that reference. In the present paper we 
do not attempt to improve the Donnachie--Landshoff fit. 
However, there appears to be a misprint in 
\cite{Donnachie:1984hf}. 
In order to reproduce a successful fit to the 
data the cutoff parameter $t_1$ for the gluon propagator 
as well as the parameter $t_0$ describing the charge 
distribution radius of the proton should be chosen as 
\be
t_0 = t_1= 0.3 \,\mbox{GeV}^2 \,,
\ee
instead of $t_0=t_1=300\,\mbox{MeV}^2$ as given 
in eq.\ (17) of the original paper \cite{Donnachie:1984hf}. 
With these changes in the original parameters we can 
reproduce the Donnachie--Landshoff fit. It is shown 
together with the relevant data in figure \ref{fig:allsdl} 
where we have chosen to restrict ourselves to the dip region 
relevant for our study. 

\begin{figure}[htbp]
\begin{center}                                                               
\input{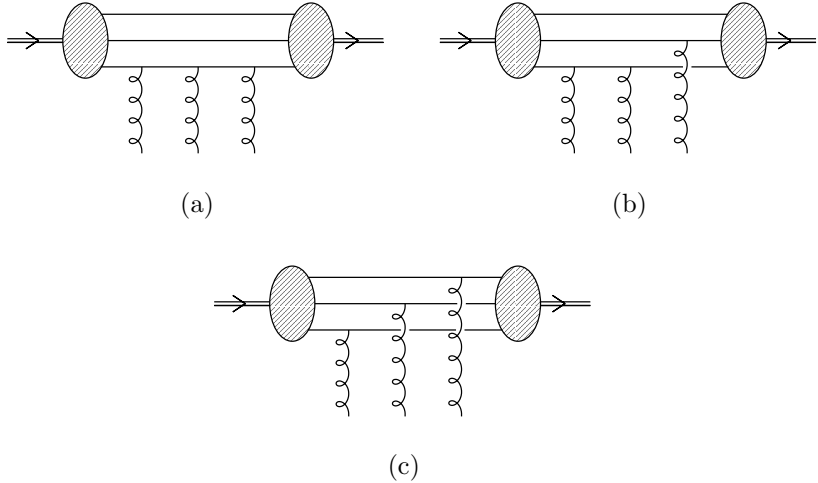}
\end{center}
\caption{Diagrams contributing to the Odderon--proton impact factor}
\label{fig:impactdiag}
\end{figure}

The Odderon contribution is particularly important 
at large $t$ and in the dip region. The dip originates 
from interference effects of the Odderon contribution 
with other contributions, in particular with those 
of Pomeron and double Pomeron exchange. 
At large $t$ the differential cross section is 
even dominated by Odderon exchange, 
leading to the observed $t^{-8}$ behaviour. 
In \cite{Donnachie:1984hf} the large-$t$ data have 
therefore been used to fix the normalization parameter 
of the Odderon contribution $T^{\cal O}$. 
In \cite{Donnachie:1979yu} it is argued that this 
dominance of the Odderon at large $t$ comes about 
because the exchange of three gluons permits to 
distribute the momentum transfer evenly between 
the three quarks in the proton. Accordingly, 
that dominant contribution corresponds to 
a situation in which each of the three gluons 
is coupled to a different quark in the proton, see 
diagram (c) in figure \ref{fig:impactdiag}. 
Also at smaller values of $t$ the authors of \cite{Donnachie:1984hf} 
use a coupling of the Odderon to the proton which is given 
only by this diagram. By selecting a single diagram only, however, 
gauge invariance is lost and the corresponding contribution 
becomes divergent as one of the gluon momenta goes to zero. 
Therefore the gluon propagator needs to be regularized by 
introducing the cut--off parameter $t_1$ (see above). 
With this procedure the coupling of the Odderon to the 
proton used in \cite{Donnachie:1984hf} leads to a 
reasonably good agreement with the data also at 
intermediate values of $t$ as shown in figure \ref{fig:allsdl}. 

A note concerning the terminology used in \cite{Donnachie:1984hf} 
seems to be in order. In that paper the Odderon contribution is 
called `three gluon exchange' but is in fact a pure $C=-1$ 
Odderon exchange. In principle, it is of course possible 
to have three gluons in a $C=+1$ state, that is in an 
antisymmetric colour state. The fact that the three gluon 
exchange in the DL fit carries 
only $C=-1$ quantum numbers is due to the particular 
coupling of the three gluons to the proton chosen in 
\cite{Donnachie:1984hf}, in which 
each of the three gluons is coupled to a different 
quark, see diagram (c) in figure \ref{fig:impactdiag}. 
This immediately implies that the three gluons are in 
a symmetric colour state. (See also equation (\ref{colorfactorc}) and 
the corresponding discussion in section \ref{highenergyinpos} below.) 
However, gauge invariance requires 
to include all possible ways of coupling the three gluons to the 
three quarks in the proton, in particular also diagrams of the 
type (a) and (b) in figure \ref{fig:impactdiag}. 

Finally, we would like to point out that the exchange of a three--gluon 
state carrying positive $C$ parity is not expected to change the DL fit. 
Due to reggeization a $C=+1$ perturbative state of three gluons has 
in the high energy limit the same analytic properties as a Pomeron 
made of two gluons \cite{Bartels:1992ym} (see also \cite{Bartels:1999aw}). 
In a fit to the data such a contribution would therefore be absorbed 
by the full Pomeron exchange and is thus effectively already contained 
in the DL fit. It is therefore not necessary to consider 
a $C=+1$ three gluon exchange separately. 

\subsection{Impact factors}
\label{impact}

In the approach using impact factors the Odderon contribution to 
the elastic proton--proton scattering amplitude is written in 
factorized form as 
\be
\label{eq:timpac}
T^{\cal O}(s,t) = \frac{s}{32} \frac{5}{6} \frac{1}{3!} \int
        \frac{d^2\delta_{1t}}{(2\pi)^4} \frac{d^2\delta_{2t}}{(2\pi)^4}
       \left[ \Phi_p(\vdelta_{1t},\vdelta_{2t},\vDelta_t)\right]^2
        \frac{1}{\vdelta_{1t}^2\, \vdelta_{2t}^2\,
                (\vDelta_t-\vdelta_{1t}-\vdelta_{2t})^2}
\,,
\ee
where the integral is over transverse momenta only. Here, 
$\vDelta_t$ is the total transverse momentum transferred in the 
$t$-channel, and $t= - \vDelta_t^2$. 
The last factor in the integral consists of the three gluon propagators 
which we assume to model the Odderon at large $t$. 
The $5/6$ originates from a colour factor and 
the $1/3!$ is implied by the exchange of three identical gluons. 
The impact factor $\Phi_p(\vdelta_{1t},\vdelta_{2t},\vdelta_{3t})$ 
describes the coupling of the Odderon to the proton; 
it is not known from first principles. 
But some of its properties can be derived from general principles. 
In order to arrive at a gauge invariant expression for the 
impact factor one needs to take into account all possible 
ways to couple the three gluons to the three quarks in the proton. 
That means one has include all three types of diagrams in 
figure \ref{fig:impactdiag} and the corresponding permutations 
of the gluon lines. 
The colour neutrality of the proton requires that the 
impact factor vanishes if one of the three transverse gluon momenta 
$\vdelta_{it}$ vanishes, 
\be
\left. 
\Phi_p(\vdelta_{1t},\vdelta_{2t},\vdelta_{3t})
\right|_{ \vdelta_{it}=0 }= 0 \,,
\:\:\:\:\:\:\:\:\:\:
\:\:\; i \in \{1,2,3\}\,, 
\ee
which ensures that potential infrared singularities 
due to the gluon propagators in the integral (\ref{eq:timpac}) 
are cancelled. 
The above property implies that the impact factor has the general form 
\be
\label{eq:impact}
\Phi_p(\vdelta_{1t},\vdelta_{2t},\vdelta_{3t})  
=  8\,(2\pi)^2\, g^3 \,\left[F(\vDelta_t,0,0) - 
                \sum_{i=1}^3 F(\vdelta_{it}, \vDelta_t-\vdelta_{it}, 0)
                +2F(\vdelta_{1t},\vdelta_{2t},\vdelta_{3t})\right]
\,,
\ee
where $\vDelta_t= \sum_{i=1}^3 \vdelta_{it}$, and 
$F(\vdelta_{1t},\vdelta_{2t},\vdelta_{3t})$ is a form factor. 
The three terms in square brackets correspond (in the 
order given in eq.\ (\ref{eq:impact})) to the 
diagram types (a), (b), and (c) in figure \ref{fig:impactdiag}, 
respectively. The form factor $F$ is related to the structure 
of the proton. Its exact form is unknown and needs to be
modelled. 

One model for the form factor $F$ was given by 
Fukugita and Kwieci\'nski in \cite{Fukugita:1979fe}, 
\be
\label{eq:formfact}
F(\vdelta_{1t},\vdelta_{2t},\vdelta_{3t}) = \frac{A^2}{A^2 + \frac12
        [(\vdelta_{1t}-\vdelta_{2t})^2 + (\vdelta_{2t}-\vdelta_{3t})^2 + 
         (\vdelta_{3t}-\vdelta_{1t})^2]}
\,.
\ee
The parameter $A$ is chosen to be half the $\rho$ mass, 
$A=384\,\mbox{MeV}$. This model for the form factor has 
recently been used in the estimate of the cross section for 
diffractive $\eta_c$ photo-- or electroproduction at HERA 
\cite{Czyzewski:1997bv,Engel:1997cg,Bartels:2001hw}.
In these references a rather large value of $\alpha_s = g^2/(4\pi) = 1$ 
has been used for the strong coupling constant. This value was 
motivated by the use of a similar value in an estimate of 
hadronic cross sections in the two--gluon model of 
\cite{Gunion:iy}. 

Another model for the form factor $F$ was proposed by Levin 
and Ryskin \cite{Levin:gg}. Their ansatz is motivated by 
a nonrelativistic quark model with oscillatory potential. 
Its explicit form is 
\be
\label{formfaclr}
F(\vdelta_{1t},\vdelta_{2t},\vdelta_{3t}) = 
\exp \left( -R_p^2 \sum_{i=1}^3 \vdelta\,^2_{it}\right)
\,.
\ee
The parameter $R_p$ is supposed to be of the order of magnitude 
of the proton radius. In \cite{Levin:gg} a value of $2.75\,\mbox{GeV}^2$ 
is given for the quantity $R_p^2$. We assume 
that the misprint is located in the exponent of the units and 
thus use $R_p^2 = 2.75\,\mbox{GeV}^{-2}$. Assuming the 
missing minus sign to be in the exponent of $R_p$ instead would 
imply an unusually small proton radius. 
The authors of \cite{Levin:gg} suggest to choose $\alpha_s=1/3$. 

\subsection{High energy scattering in position space}
\label{highenergyinpos}
In this subsection we give a very short recapitulation of the basic
ideas of the treatment of high energy scattering 
developed by Nachtmann \cite{Nachtmann:1991ua}, for details 
and further justification we refer the reader to the original article. 
The method is based on the functional representation of scattering
matrix elements and the WKB method. 
We first consider quark--quark scattering amplitudes in an 
external colour field using the WKB approximation. The 
quantization is done by functional integration. 
Nucleon--nucleon scattering amplitudes are obtained 
from scattering amplitudes of clusters of quarks by averaging 
over wave functions in transverse space. This is an alternative 
to the treatment of high energy scattering in momentum 
space and particularly suited for investigating the 
effects of the spatial structure of the hadrons. In the present 
paper we use perturbative three--gluon exchange to model 
the Odderon, but the method presented here also allows one 
to easily incorporate nonperturbative models. 

The first step is to transform the $S$-matrix element of
two quarks with incoming momenta $p_1,p_2$ and outgoing momenta
$p_3,p_4$ into a Green function. 
This is done by means of the LSZ reduction formalism, 
\bea
\label{ha1}
\langle\, p_3\,p_4^{\rm out}|p_1\,p_2^{\rm in}\,\rangle 
&=& Z_\ps^{-2} \int d^4x_1\cdots d^4x_4
\exp \left[i(p_3x_3+p_4x_4-p_1x_1-p_2x_2)\right] \times
\nn \\
&& \times \langle\, {\rm T} \bar u(p_3) f(x_3)\bar u(p_4) f(x_4)
\bar f(x_1) u(p_1) \bar f(x_2) u(p_2)\,\rangle\,, 
\eea
where $f(x) = (i\ga\pa - m) \ps(x)$ and 
$Z_\psi$ is the wave function renormalization. 

Next, the four point function 
$\langle\, {\rm T} \ps(x_3)\ps(x_4)\bar \ps(x_1)\bar \ps(x_2)\,\rangle$
contained in the rhs of (\rf{ha1})
is expressed as a functional integral over the quark 
and the gluon fields, $\psi$ and $B$ respectively, 
written formally as 
\beq
\langle\,{\rm T} \ps(x_3)\ps(x_4)\bar \ps(x_1)\bar \ps(x_2)\,\rangle = 
\int \cD\ps\,
\cD \bar \ps \cD B \,\ps(x_3)\ps(x_4)\bar \ps(x_1)\bar \ps(x_2)
\exp[-i S_{\rm full QCD}]\,, \lbq{ha2}
where $S_{\rm full QCD}$ is the full QCD action.
The fermion integration is Gaussian and can therefore be performed, yielding 
\beqa \lb{ha3}
&&\hspace{-3mm}
\langle\,{\rm T} \ps(x_3)\ps(x_4)\bar \ps(x_1)\bar \ps(x_2)\,\rangle =
\int \cD B \det[-i(i\ga D - m)]\times\\
&& \times \left[ S_F(x_3,x_1;B)\,S_F(x_4,x_2;B)+
S_F(x_3,x_2;B)\,S_F(x_4,x_1;B) \right]
\exp[-iS_{\rm pure QCD}] \,,
\nn \enqa
which contains the functional determinant of the Dirac operator,
and the quark propagators $S_F(x_i,x_j;B)$ 
in the external colour potential $B_\mu^F$. 
The functional integration is now to be 
performed only over the gluon fields with the pure QCD action (i.\,e.\ without
quark contribution) in the measure. If we concentrate on lowest order
exchange  the determinant 
can be set to one. Furthermore, if we are interested only in  momentum
transfer small compared 
to the total energy the second ($u$-channel) term in 
the integrand can be neglected. Collecting all factors we finally obtain 
\beq
\langle\, p_3\,p_4^{\rm out}|p_1\,p_2^{\rm in} \,\rangle
\approx Z_\ps^{-2}\, \int \cD B\,
\cS(p_3,p_1;B) \cS(p_4,p_2;B) \exp[-i S_{\rm pure QCD}] \,,
\lbq{ha4} 
where $\cS(p_i,p_j;B)$ is the scattering matrix element of a quark with
momentum $p_j$ to one with momentum $p_i$ in an external colour field $B$.

In the next step we have to find a suitable form for the $S$-matrix element 
$\cS(p_i,p_j;B)$. 
One can show \cite{Nachtmann:1991ua} that the quark scattering
matrix elements $\cS(p_i,p_j;B)$ in an external field
can be expressed as a generalized WKB expression 
\beq
\cS(p_i,p_j;B) = \bar u(p_i) \ga^\mu u(p_j) 
{\rm P}\exp\left[-ig\int_\Ga \bB_\rh\,dx^\rh\right]
\left(1+O\left(\frac{1}{p^0_i}\right)\right)\,,
\lbq{ha5}
where we denote by $\bB$ the Lie--algebra valued gauge potential. The
path--ordered integral is taken along the classical path $\Gamma$. From 
the scattering amplitudes for single quarks in the gluon field 
we obtain, according to (\rf{ha4}), the nonperturbative quark--quark scattering
amplitude by integrating  the  product of the two scattering amplitudes over
the gluon field. More specifically, 
consider two quarks travelling along the light--like paths $\Gamma_1$ and
$\Gamma_2$ given by 
\begin{equation}
\Gamma_1=(x^0,\vec b/2, x^3=x^0) \quad \mbox{ and }\quad \Gamma_2=
(x^0, -\vec b/2, x^3= -x^0)  , 
\label{3A2}
\end{equation}
corresponding to quarks moving in opposite directions 
with the velocity of light, with an impact vector $\vec b$ in the $x^1 x^2$-plane
(referred to in the following as the transverse plane). 
Let $\bV_{i}(\pm\vec b/2)$ be the phases picked up by the 
quarks along these paths, 
\begin{equation}
\bV_{i}(\pm\vec b/2)={\rm P}\exp\left[-ig \int_{\Gamma_{i}} {\bf B}_\mu
(z)\ dz^\mu\right]~. 
\label{3A3}
\end{equation}
Then the $S$-matrix element for two quarks with momenta $p_1$, $p_2$
and colour indices $\al_1$, $\al_2$ leading to two quarks of 
momenta $p_3$, $p_4$ and
colours  $\al_3$, $\al_4$ is \cite{Nachtmann:1991ua}
\begin{equation}
S_{\al_3\al_4;\al_1\al_2}(s,t)=\bar{u} (p_3) \gamma^\mu u(p_1) \bar{u} (p_4)
\gamma_\mu u(p_2)\,{\cal V}\,, 
\label{3A4}
\end{equation}
where 
\begin{equation}
{\cal V}=i Z^{-2}_\psi 
\left\langle\int d^2 b\ e^{-i\vec q \cdot \vec b}
\left[ {\bf V}_1\left(-{\vec b\over 2}\right)\right]_{\al_3\al_1}
\left[{\bf V}_2 \left(+{\vec b\over 2}\right)\right]_{\al_4\al_2}
\right\rangle\,. 
\label{3A5}
\end{equation}
Here $\langle \,\cdot\,\rangle$ denotes functional integration over the
gluon field, and $\vec q$ is the momentum transfer $(p_1-p_3)$ projected
onto the transverse plane. Of course the approximation makes sense only  
if $|\vec q\,|\ll|\vec p\,|$. 

In the limit of high energies we have  helicity conservation, 
\begin{equation}
\bar u(p_3)\ \gamma^\mu u(p_1)\ \bar u(p_4)\ \gamma_\mu u(p_2)
\:\:\:\:
\mathop{\longrightarrow}_{s \to \infty}
\:\:\:\:
2s\delta_{\lambda_3\lambda_1}
\delta_{\lambda_4 \lambda_2} \,, 
\label{3A7}
\end{equation}
where $\lambda_i$ are the helicities of the quarks and $s=(p_1+p_2)^2$.
In the following we can thus ignore the spin degrees of freedom. 

In order to come to the nucleon--nucleon scattering amplitude 
we first consider the scattering of two groups of three quarks 
moving on parallel lightlike world lines, each of which has the form
\beq
\hat{\Ga}^a_1(x_0,\vec b/2+\vec x\,_1^a, x^3=x^0),~~
\hat{\Ga}^a_2(x_0,-\vec b/2+\vec x\,_2^a, x^3=-x^0),~~ a=1,2,3 \,.
\enq
In order to ensure that these quark clusters asymptotically form 
colour singlet states 
all colours are parallel--transported in the remote past and future to a
reference point of the cluster and there contracted antisymmetrically.
This leads to the following $S$-matrix element \cite{Dosch:1994ym} 
for scattering of colour--neutral clusters, 
\bea
\lb{nuc1}
\lefteqn{S\left(\vec{x}\,_1^1,\vec{x}\,_1^2,\vec{x}\,_1^3,
\vec{x}\,_2^1,\vec{x}\,_2^2,\vec{x}\,_2^3\right)= 
\frac{1}{36}\frac{1}{Z_1 Z_2} \times} 
\\
&&
\times \left\langle\ep_{\al\beta\ga}
\left({\bf V}^1_1\right)_{\al\al'}\left({\bf V}^2_1\right)_{\beta\beta'}
\left({\bf V}^3_1\right)_{\ga\ga'}\ep_{\al'\beta'\ga'}
\ep_{\rh\mu\nu}\left({\bf V}^1_2\right)_{\rh\rh'}
\left({\bf V}^2_2\right)_{\mu\mu'}
\left({\bf V}^3_2\right)_{\nu\nu'}\ep_{\rh'\mu'\nu'}\right\rangle \,. 
\nn
\eea 
The non-Abelian phase factors $\bV_i^a$ are defined as 
in (\rf{3A3}) with the $\sqcup$-shaped integration paths $\Gamma_i$ 
as indicated in figure \rf{baryon} for one cluster. 
\begin{figure}[htb]
\begin{center}
\epsfysize 6cm
\epsfbox{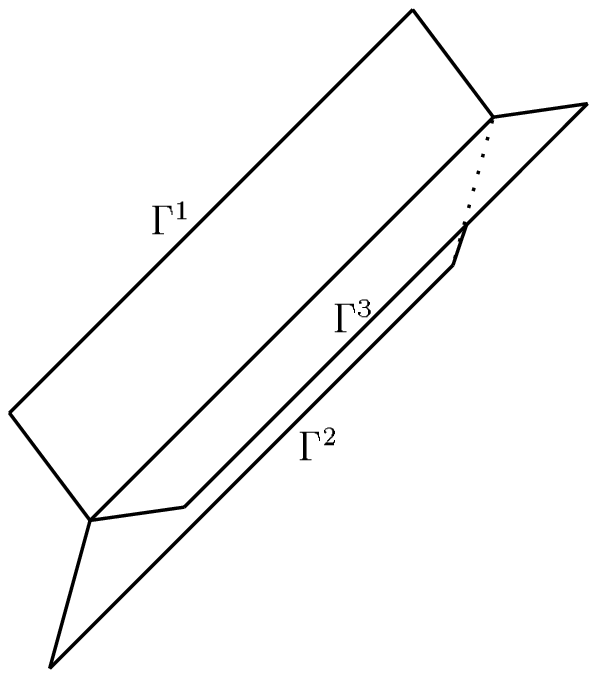}
\end{center}
\caption{The paths in a colour neutral three--quark cluster}
\label{baryon}
\end{figure}                                                                 
The $Z_i$ denote again the wave function renormalization for the 
respective clusters which in lowest order can be set equal to one. 
We introduce the reduced scattering amplitude $J$ related 
to the $S$-matrix element for the scattering of quark clusters, 
\beq
\label{JisSminus1}
J(\vec x\,_1^1,\vec x\,_1^2,\vec x\,_1^3,
\vec x\,_2^1,\vec x\,_2^2,\vec x\,_2^3)= 
S(\vec x\,_1^1,\vec x\,_1^2,\vec x\,_1^3,\vec x\,_2^1,
\vec x\,_2^2,\vec x\,_2^3)- 1 \,.
\enq

The differential nucleon--nucleon cross section is obtained from the gauge
invariant scattering amplitude $T(s,t)$ via (\ref{dsigdtfromT}). 
The Odderon contribution $T^{\cal O}(s,t)$ to the 
scattering amplitude $T(s,t)$ is computed by integrating 
over the  transverse coordinates with a suitable transverse
wave function $\psi$, 
\be
T^{\cal O}(s,t)= 2i\, s \int d^2 b \,e^{-i \vec q \cdot \vec b} \int 
d^6{\mathbf{ R}}_1 
d^6{\mathbf{ R}}_2 
|\psi({\mathbf{ R}_1})|^2
|\psi({\mathbf{ R}_2})|^2 
J(\vec x\,_{1}^{1},\vec x\,_{1}^{2},\vec x\,_{1}^{3},\vec x\,_{2}^{1},
\vec x\,_{2}^{2},\vec x\,_{2}^{3})  \,,
\ee
where the colour indices in (\ref{JisSminus1}) are symmetric, and 
${\mathbf{ R}}_i$ denotes the set of positions of the quarks relative 
to the centre of nucleon~$i$,
and $\vec b$ is the impact vector between the two nucleons 
(see figure \rf{twostar}), 
\begin{figure}[htb]
\begin{center}
\epsfysize 3.5cm
\epsfbox{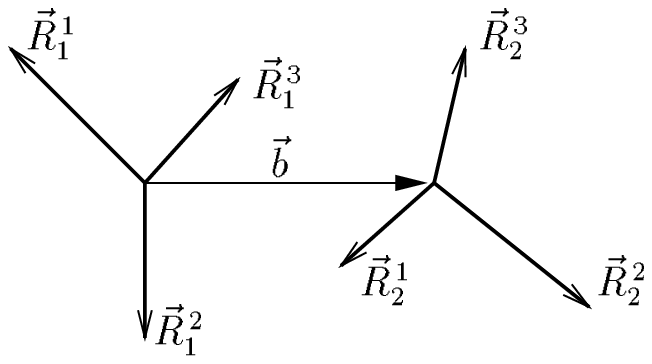}
\end{center}
\caption{The positions of the quarks in transverse space}
\label{twostar}
\end{figure}
\beq
\mathbf{ R}_i=\left(\vec R\,_i^1,\vec R\,_i^2,\vec R\,_i^3\right)\,, \:\:\:\:\:
\vec x\,_1^a = \frac{\vec b}2 + \vec R\,_1^a\,, \:\:\:\:\:
\vec x\,_2^a= -\frac{\vec b}2 +\vec R\,_2^a \,.
\enq

For a perturbative evaluation of
three gluon exchange we expand ${\bf V}^a_i$ in (\ref{nuc1}) 
up  to order $g^3$.
Expanding the matrix valued field in generators $\ta$  of $su(3)$, 
\beq
\label{integexpa}
\int_{\Ga^a_i}dz^\mu~\bB_\mu(z)= \hat B_{a,i}^c \ta^c \,,
\enq
we obtain 
\beqa
\label{viexpand}
\left(\bV^a_i\right)_{\al\beta} &=&
\de_{\al\beta} - i g  \hat B_{a,i}^c \ta^c_{\al\beta}
-\frac{1}{2} g^2 
\hat B_{a,i}^c\hat B_{a,i}^{c'}(\ta^c\ta^{c'})_{\al\beta} \nn \\
&&-\frac{i^3}{3!} g^3 
\hat B_{a,i}^c\hat B_{a,i}^{c'}\hat B_{a,i}^{c''}
(\ta^c\ta^{c'}\ta^{c''})_{\al\beta} + {\cal O}(g^4) \,.
\enqa
In principle we have to take into account path ordering in the 
integral (\ref{integexpa}). But since we are only interested in 
the Odderon contribution, which is symmetric in the colour 
indices, we may discard it here. To that order we also do not need to 
take into account the parallel transporters from the reference 
points to the light-like paths and we have set $Z_i =1$. 

The lowest order three gluon exchange contribution to (\rf{nuc1}) is obtained
by pairing three fields of group (1) with three of group (2).
In each group we have three possibilities:\\
a) two quarks are not involved\\
b) one quark is not involved\\
c) all quarks are involved.\\
We consider first the respective colour tensors 
$C^{\rm a}$, $C^{\rm b}$, $C^{\rm c}$. 
For case a) we have 
\beq
C_{c c' c''}^{\rm a}=
\ep_{\al\beta\ga}\ep_{\al'\beta'\ga'}\de_{\al\al'}\de_{\beta\beta'}
(\ta^c\ta^{c'}\ta^{c''})_{\ga\ga'}
=\frac{1}{2}d_{cc'c''} + \frac{i}{2}f_{cc'c''}
\,,
\enq
where $f$ are the structure constants of $su(3)$ and $d$ the symmetric 
constants occurring in the anti--commutators. 
For case b) we obtain in the same way the colour factor
\beq
C_{c c' c''}^{\rm b}=-\frac{1}{4}d_{cc'c''} - \frac{i}{4}f_{cc'c''}
\,,
\enq
and for the case where all quarks in the nucleon occur in  pairings
\beq
C_{c c' c''}^{\rm c} =\frac{1}{2}d_{cc'c''} \,.
\label{colorfactorc}
\enq

Only the symmetrically coupled colours  contribute to $C=-1$ exchange.
Since in the treatment of DL only case c) was considered they 
automatically had only a negative charge parity contribution 
(see above). 
We are only interested in the $C=-1$ contribution
and because of the symmetry the path ordering has no influence. 
This simplifies the calculation considerably, since now we can perform the
integrations along the lightlike paths without restriction and this
leads to a projection into the transverse space. 
The general structure of a contribution to (\rf{nuc1}) is therefore given by
the product of two colour factors $C$ given above and the product of three
gluon propagators in transverse space connecting a quark of group (1) 
with one of group (2).

We therefore obtain for the reduced scattering amplitude 
\beqa
\label{contractionsum}
\lefteqn{J(\vec x\,_{1}^{1},\vec x\,_{1}^{2},\vec x\,_{1}^{3};
\vec x\,_{2}^{1},\vec x\,_{2}^{2},\vec x\,_{2}^{3})=}\nonumber \\
&& g^6 
\sum_{a_i,b_i=1}^3 K(a_1,a_2,a_3;b_1,b_2,b_3)
\chi(\vec x\,_1^{a_1},\vec x\,_2^{b_1}) \chi(\vec x\,_1^{a_2},\vec x\,_2^{b_2})
\chi(\vec x\,_1^{a_3},\vec x\,_2^{b_3}) \,.
\enqa
The factor $K$ is obtained from (\ref{nuc1}), 
(\ref{JisSminus1}) and (\ref{viexpand}) taking into account 
only colour symmetric gluon states. 
The corresponding contractions and the combinatorial factors 
arising in the sum over indices in (\ref{contractionsum}) 
have been calculated in \cite{Rueter:1996yb,Rueterthesis}. 
We refer the reader to \cite{Rueterthesis} for the somewhat 
cumbersome details. 
$\chi$ is the gluon propagator in transverse space, 
\beqa
\chi(\vec x,\vec y)&=&  \int \frac {d^2k}{(2 \pi)^2} \frac{1}{\vec
k\,^2+m^2} e^{-i \vec k \cdot (\vec x-\vec y)}\\
&=& \frac{1}{2 \pi} K_0\left( m \left|\vec x-\vec y \right|\right)
\,,
\enqa
where $K_0$ is the modified Bessel function. 
The single diagrams are infrared divergent. In order to 
regularise them we have introduced a gluon mass $m$ which 
is possible in LO approximation. In the final gauge 
invariant expressions we can set the gluon mass to zero. 

\begin{figure}[htb]
\begin{center}
\epsfysize 3.5cm
\epsfbox{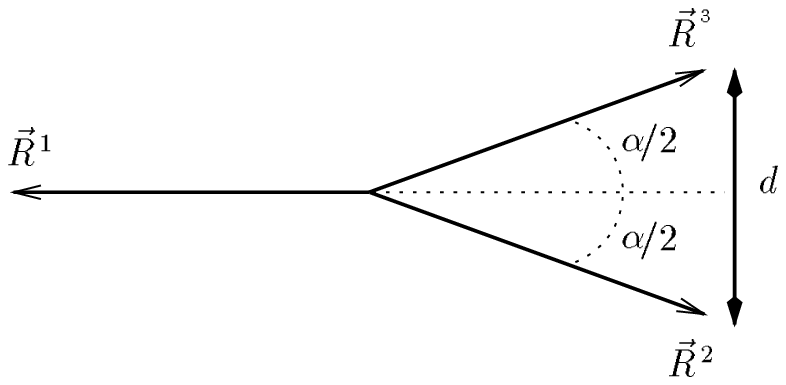}
\end{center}
\caption{Definition of the angle $\alpha$ characterising the 
proton configuration}
\label{star}
\end{figure}

For the quark density in the nucleon we make the 
simple ansatz 
\begin{equation}
\label{wavefunction}
\left| \psi(\vec R_1,\vec R_2,\vec R_3)\right|^2=
\frac2\pi \frac1{S_p^2} \,\exp\left(-\frac{2R_1^2}{S_p^2}\right)\,
\delta^2(\vec R_2 - {\bf M}_\beta\,\vec R_1)\,
\delta^2(\vec R_3 - {\bf M}_{-\beta}\,\vec R_1)\,,
\end{equation}
where ${\bf M}_\beta= \pmatrix{\cos\beta&-\sin\beta\cr\sin\beta&\cos\beta\cr}$
and $\beta=\pi-\alpha/2$. 
The quantity $S_p$ determines the electromagnetic radius of the
nucleon. We choose $S_p= 0.8\,\mbox{fm}$ in the range given in  
\cite{Rueterthesis,Dosch:2001jg}. 
The meaning of the angle $\alpha$ is illustrated in figure \ref{star}. 
The value $\alpha=\frac{2 \pi}{3}$ corresponds to a Mercedes
star configuration of the quarks in the nucleon. 
$\alpha=0$ corresponds to a quark--diquark picture of 
the nucleon with an exactly pointlike diquark. 
For small angles $\alpha$ we can still speak of a diquark--cluster 
in the nucleon, and we call the distance $d$ between the two 
quarks in such a cluster the diquark size, see figure \ref{star}. 
With the wave function (\ref{wavefunction}) 
one then obtains for the average diquark size $\langle d \rangle$ 
\be
 \langle d \rangle = \sqrt{\frac{\pi}{2}} \, 
\sin \left(\frac{\alpha}{2} \right)  \, S_p \,.
\ee

\section{Results}
\label{resultssec}

In the following we use the Donnachie--Landshoff fit as a 
framework for confronting different models for the 
coupling of the Odderon to the proton with the $pp$ and 
$p\bar{p}$ elastic scattering data in the dip region. 
This is done by replacing the Odderon contribution 
$T^{\cal O}$ to the scattering amplitude $T$ in the DL fit 
by the other models for $T^{\cal O}$ 
discussed in sections \ref{impact} and \ref{highenergyinpos}. 
The integrations in the calculation of the differential cross 
section are performed numerically. 

\begin{figure}[phtb]
\def\size{\normalsize}
\begin{center}                                                               
\input{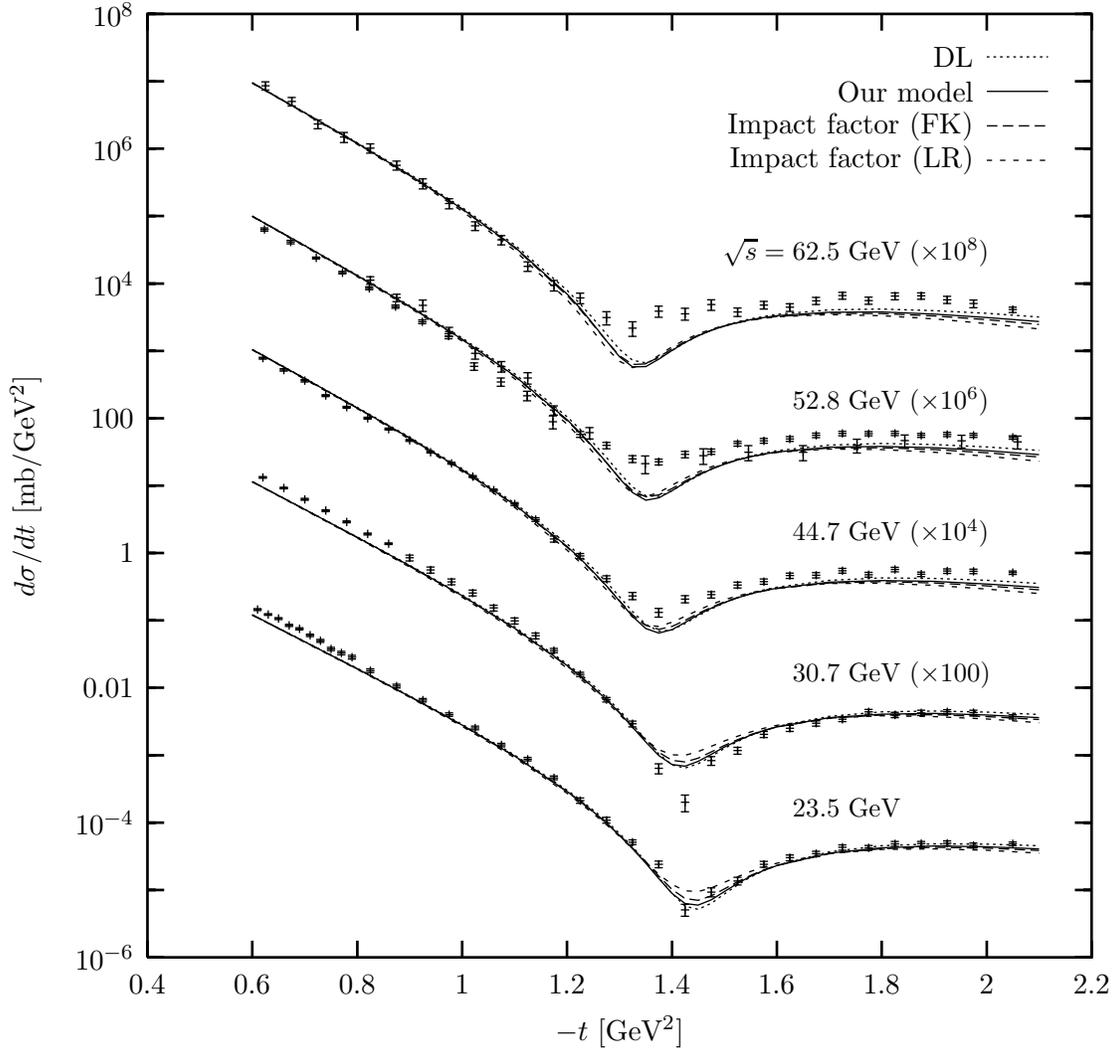}
\end{center}
\caption{Differential cross section for elastic $pp$ scattering 
calculated using different couplings of the Odderon to the proton: 
the original Donnachie--Landshoff fit (dotted), our geometrical 
model for the proton (solid), 
and the Fukugita--Kwieci\'nski (FK, long--dashed) 
and Levin--Ryskin (LR, short--dashed) impact factors}
\label{fig:allcurves}
\end{figure}

The results for the differential cross section in the dip region 
are shown in figure \ref{fig:allcurves} together with the 
relevant data. 
For comparison we also show the Donnachie--Landshoff fit 
described already in section \ref{dlsection} as the 
dotted line in this figure. 

\begin{figure}[htb]
\begin{center}
\input{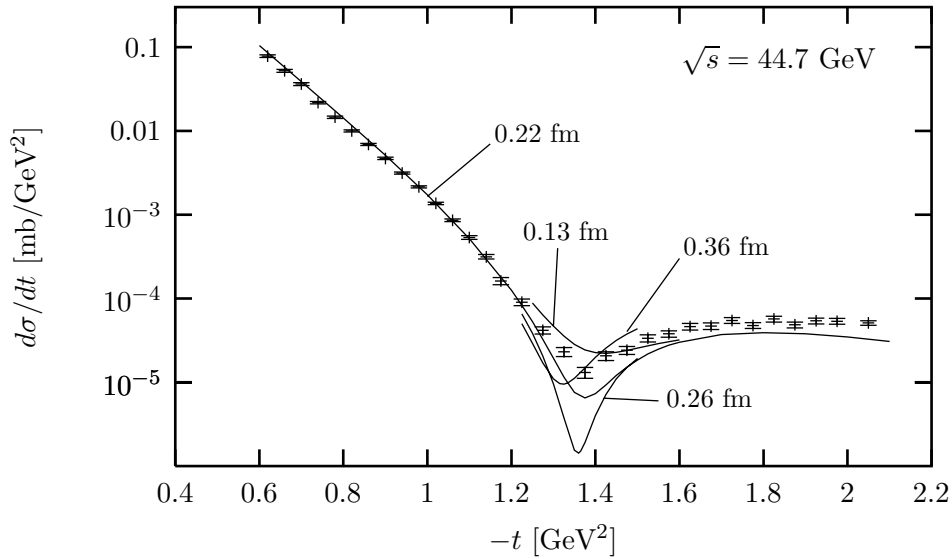}
\end{center}
\caption{Dependence of the differential cross section on the average 
diquark size $\langle d \rangle$ chosen in our geometric 
model of the proton for fixed coupling constant $\alpha_s=0.4$}
\label{fig:angles}
\end{figure}

The solid line in figure \ref{fig:allcurves} represents the result 
obtained with the geometric model for the proton described 
in section \ref{highenergyinpos}. It almost coincides with 
the DL fit and gives a satisfactory description of all available data. 
We have fixed the value of 
the strong coupling at $\alpha_s=0.4$ and then adjusted the 
angle $\alpha$ characterising the proton configuration. 
The optimal description of the data is obtained for $\alpha =  0.14 \,\pi$, 
corresponding to an average diquark size of $0.22\,\mbox{fm}$. 
For other choices of the average diquark size (or equivalently of 
the angle $\alpha$) and fixed $\alpha_s=0.4$ 
the description of the data becomes much 
worse as is illustrated in figure \ref{fig:angles} for one 
centre--of--mass energy, $\sqrt{s} = 44.7\,\mbox{GeV}$. 
The situation is very similar for the other centre--of--mass 
energies. With increasing average diquark size $\langle d \rangle$ 
the minimum of the differential cross section moves towards  smaller $t$. 
At the same time the depth of the dip changes in such a way 
that the optimal value of $\langle d \rangle$ can 
be determined with only a small uncertainty. 

The parameters $\alpha_s$, $\alpha$, and $S_p$ are of course 
strongly correlated in their effect on the differential cross section. 
Since $S_p$ is rather strictly constrained by the electromagnetic 
size of the nucleon we do not vary it here. The constraints on the 
other two parameters in our model, on the other hand, are only 
weak. The correct value of the strong coupling constant $\alpha_s$ 
is not known precisely in the dip region but has a strong effect 
on the cross section as it enters in the third power on the 
amplitude level already. The correct value for the angle $\alpha$ is 
even less constrained, and also the variation of $\alpha$ has a 
strong effect on the cross section. This is particularly true for small values 
of $\alpha$ (or small diquark sizes) which are known to imply 
a strong suppression of the amplitude. In the framework of our 
present investigation it is obviously not possible to determine 
$\alpha_s$ and the angle $\alpha$ independently. 
We have therefore determined the optimal value for $\alpha$ 
also for other choices of $\alpha_s$ than the one mentioned 
above. For the choice $\alpha_s=0.3$, for instance, we find that 
the best description of the data results for $\alpha= 0.22\,\pi$, 
corresponding to an average diquark size of 
$\langle d \rangle = 0.34\,\mbox{fm}$. Choosing $\alpha_s=0.5$ 
instead, we find an optimal value of $\alpha= 0.095\,\pi$, 
corresponding to $\langle d \rangle = 0.15\,\mbox{fm}$. 
We would like to point out that the resulting sizes of the 
diquark cluster in the nucleon are thus rather small for all 
reasonable choices of $\alpha_s$ at the relevant momentum 
scale in the dip region.  
A Mercedes star configuration in the proton would in fact 
imply an unrealistically small value of $\alpha_s \simeq 0.17$. 
This result of course assumes that LO perturbation theory 
can be applied in the dip region. 

\begin{figure}[htb]
\begin{center}
\input{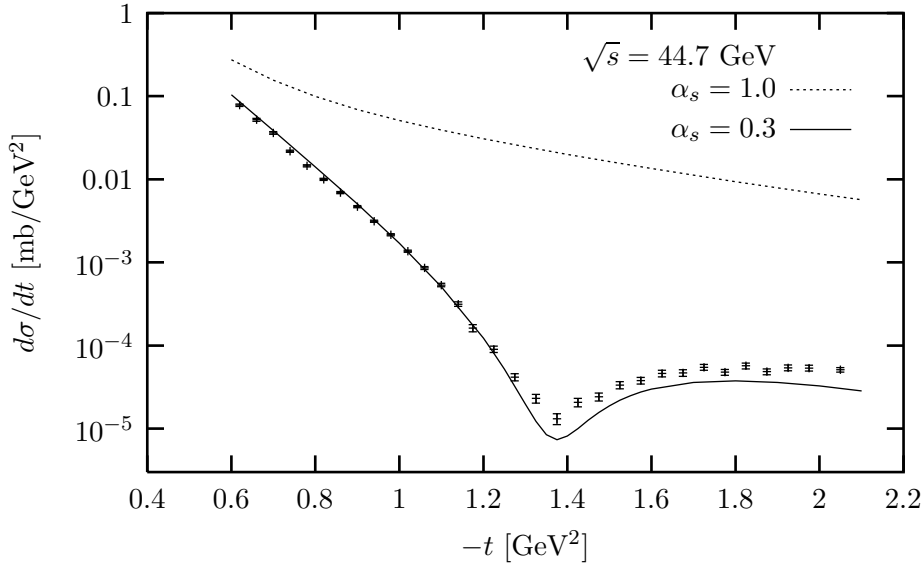}
\end{center}
\caption{Dependence of the differential cross section 
obtained from the Fukugita--Kwieci\'nski impact factor 
on the choice of $\alpha_s$}
\label{fig:impactdiffalpha}
\end{figure}

We now turn to the models for the Odderon--proton 
impact factor described in section \ref{impact}. 
Both models contain two parameters one of which is the strong 
coupling $\alpha_s$. The other one is 
in the case of the Fukugita--Kwieci\'nski (FK) 
model the parameter $A=m_\rho /2$, in the case of the 
Levin--Ryskin (LR) model it is the parameter $R_p$. 
The latter parameters are again related to the proton 
size and should thus be considered strongly constrained. 
We therefore keep them at the values given in the original 
papers (as quoted in section \ref{impact}) and vary only $\alpha_s$. 
The differential cross section obtained with the 
Fukugita--Kwieci\'nski model (\ref{eq:formfact}) 
for the impact factor is 
shown as the long--dashed curve in figure \ref{fig:allcurves}. 
It gives an equally good description of the data as the DL fit 
and as our geometric model of the proton. 
In order to obtain this curve we have chosen $\alpha_s=0.3$ 
instead of the value $\alpha_s=1.0$ originally proposed in 
\cite{Fukugita:1979fe}. Had we chosen that value instead, 
the resulting differential cross section would dramatically overshoot 
the data and not even show a dip structure, 
as is illustrated for one centre--of--mass energy 
($\sqrt{s}=44.7\,\mbox{GeV}$) in figure \ref{fig:impactdiffalpha}. 
Also the Levin--Ryskin model (\ref{formfaclr}) 
for the impact factor leads to a good description of the 
data when the strong coupling constant is chosen as 
$\alpha_s=0.5$. The corresponding differential cross 
section is shown as the short--dashed curve in figure 
\ref{fig:allcurves}. Also here the dependence of the cross 
section on $\alpha_s$ is very strong, actually being the 
same as in the case of the FK model 
as can be easily seen from eq.\ (\ref{eq:impact}). 

Finally, we turn to the differential cross section for 
elastic $p\bar{p}$ scattering. Unfortunately, sufficiently 
many data points are available only for one centre--of--mass 
energy in the ISR range, $\sqrt{s} = 53\,\mbox{GeV}$. 
Our results for that energy are shown in figure \ref{fig:ppbar}. 
\begin{figure}[htb]
\begin{center}
\input{dsigdt59_ppbar53.ptex}
\end{center}
\caption{Differential cross section for elastic $p\bar{p}$ scattering 
at $\sqrt{s} = 53\,\mbox{GeV}$ as 
calculated using different couplings of the Odderon to the proton: 
the original Donnachie--Landshoff fit (dotted), 
our geometrical model for the proton (solid), 
and the Fukugita--Kwieci\'nski (FK, long--dashed) 
and Levin--Ryskin (LR, short--dashed) impact factors, 
data taken from \protect\cite{Breakstone:1985pe}}
\label{fig:ppbar}
\end{figure}
Here we have used the same parameters as for the curves in 
figure \ref{fig:allcurves}. Again, our geometric model as well 
as the two models for the impact factors lead to a description 
of the data which is as good as the Donnachie--Landshoff fit, 
producing a shoulder rather than the dip observed in $pp$ scattering. 

In summary we can say that the experimental data available 
in the dip region are by far not precise enough to distinguish between 
different models for the coupling of the Odderon to the proton. 
All models for that coupling and the corresponding models 
for the proton structure lead to a satisfactory description of the 
data when the respective parameters are chosen appropriately. 
But for a given model these parameters are quite strongly 
constrained by the data. This applies in particular to the value 
of $\alpha_s$ in the two models using impact factors. 

\section{Conclusions}
\label{conclusions}

The only clear experimental evidence for the existence of an Odderon 
comes from measurements of the differential cross section 
for high energy elastic $pp$ and $p\bar{p}$ scattering in the 
dip region at around $|t| \simeq 1.3\, \mbox{GeV}^2$. 
The Odderon contribution to this process is expected to be sensitive 
to the proton structure. In the present paper we have studied 
different models for the Odderon--proton coupling. 
As a framework we have used the Donnachie--Landshoff fit 
which successfully describes all available data for this process, and 
we have replaced the Odderon contribution to this fit by the respective 
model. We have taken two models for the Odderon--proton coupling 
from the literature. These two models are based  on 
impact factors in momentum space. In addition, we have constructed  
a geometric model for the proton in which the effect of a possible 
diquark cluster can be studied. In all three cases the Odderon 
is modelled by perturbative three--gluon exchange in the 
$C=-1$ channel. 

We find that all models for the Odderon--proton coupling give 
very similar results if the model parameters, in particular the 
strong coupling constant, are chosen appropriately. All models work 
as well as the original Donnachie--Landshoff fit. 
The available data cannot 
distinguish between the different models. But for a given model 
the data impose very strong constraints on the parameters of 
that model. Using our geometric model we find that the average 
size of the diquark cluster in the proton is quite small, 
$\langle d \rangle < 0.35\,\mbox{fm}$. This result is obtained when 
assuming that reasonable values for strong coupling constant 
$\alpha_s$ in the dip region are larger than $0.3$. 
In the nonperturbative model used in \cite{Rueter:1996yb} 
such a small diquark is sufficient to explain the absence 
of an Odderon signal in the ratio of the real to imaginary 
part in the forward direction \cite{Augier:1993sz}. This 
can be understood easily. In the nonperturbative model for 
the IR behaviour of QCD soft gluons dominate and therefore 
the resolution is much coarser. 

It turns out that in the models based on Odderon--proton impact factors 
the data impose rather strong constraints on the choice of the 
strong coupling constant $\alpha_s$ which appears as a parameter 
in these models. In the case of the impact factor proposed by Levin 
and Ryskin we find that $\alpha_s$ has to be chosen as $0.5$, 
i.\,e.\ a value rather close to the $1/3$ proposed originally. 

Of particular interest is the model for the Odderon--proton 
impact factor proposed by Fukugita and Kwieci\'nski. 
Recently, this model has been used for the calculation of 
different processes, among them the diffractive photo-- and 
electroproduction of $\eta_c$ mesons at HERA. This process 
is currently considered to be one of the best possible ways to 
observe the Odderon experimentally. The corresponding calculations 
\cite{Czyzewski:1997bv,Engel:1997cg,Bartels:2001hw} 
use a rather large value $\alpha_s=1$ in the impact factor. 
In order to describe the data for $pp$ elastic scattering with this 
impact factor, however, 
we find that $\alpha_s$ needs to be chosen as $0.3$. 
This observation indicates that the current estimates 
for diffractive $\eta_c$ production at HERA might be somewhat 
optimistic. 

In our study we have assumed that the Odderon can be modelled 
by perturbative three--gluon exchange. However, the dip region 
of $pp$ elastic scattering is located at momentum transfers $\sqrt{t}$ 
just slightly above $1\,\mbox{GeV}$, that is at the lowest edge 
of the applicability of perturbation theory. 
It would therefore be very desirable to study this process also in the 
framework of a nonperturbative model. 

\section*{Acknowledgements}
We would like to thank J.\ Kwieci\'nski and O.\ Nachtmann 
for helpful discussions.

\end{document}